\documentclass{jetpl}

\twocolumn

\lat

\DeclareMathOperator{\Tr}{Tr}

\DeclareMathOperator{\ep}{\epsilon}

\DeclareMathOperator{\h}{\hbar }  
  
\DeclareMathOperator{\vro}{\varrho}

\DeclareMathOperator{\pa}{\partial} 
\DeclareMathOperator{\pr}{^{\prime}}

\DeclareMathOperator{\opi}{\widehat{\Pi}}

\DeclareMathOperator{\mcn}{\mathcal{N}}

\DeclareMathOperator{\df}{\mathcal{D}}
\DeclareMathOperator{\ef}{\mathcal{E}}

\DeclareMathOperator{\jof}{\widehat{\mathcal{J}}}

\DeclareMathOperator{\lof}{\widehat{\mathcal{L}}}

\DeclareMathOperator{\qof}{\widehat{\mathcal{Q}}}

\DeclareMathOperator{\dtf}{\widetilde{\mathcal{D}}}

\DeclareMathOperator{\qtf}{\widetilde{\mathcal{Q}}}

\DeclareMathOperator{\vtf}{\widetilde{\mathcal{V}}}

\title{Time-energy uncertainty as cause of thermal flicker noise}

\rtitle{Time-energy uncertainty and flicker noise}

\sodtitle{Time-energy uncertainty as cause of thermal flicker noise}
 
\author{Yu.\,E.\,Kuzovlev\/\thanks{kuzovlev@kinetic.ac.donetsk.ua, 
yuk-137@yandex.ru}}

\rauthor{Yu.\,E.\,Kuzovlev}

\sodauthor{Yu.\,E.\,Kuzovlev}

\address{Donetsk Free Statistical Physics Laboratory}

\abstract{It is shown that 
if kinetics of quantum transitions 
takes account of energy uncertainty of intermediate states, 
then it creates  
non-decaying correlations and 
non-averagable (flicker) fluctuations 
in the energy as well as 
in rates of transitions-induced irreversible processes, 
in particular, 
flicker noise or maybe suppression 
of mobility (rate of wandering)
of particle interacting with 
thermally equilibrium medium.}


\begin{document}

\maketitle

1.\, Effects of interactions in systems of many particles (degrees of freedom) 
usually are thought in terms of 
such random interaction events as quantum transitions.  
Then one has to prescribe them 
unambiguously definite personal probabilities.
But this is impossible without applying of violence 
to exact equations of statistical mechanics.  
Frequently the violence is bringing in 
the non-stationary perturbation theory (NPT) \cite{ll3} 
so called ``Fermi golden rule'' (GR) \cite{vk,hj}, 
that is replacement  
\[  
\frac {2\,[ 1 -\cos\,(\tau\ef_{21}/\h)]} { \ef_{21}^2} \, |\Phi_{21} |^2 
\, \Rightarrow\, \frac {2\pi\tau}\h \, |\Phi_{21} |^2 \,\delta(\ef_{21}) \,\, . \,\,  
\] 
Here, for a system with Hamiltonian \,$H=  H_0 + \Phi$\,, 
  \,$\Phi_{21}$\, are matrix elements  (ME) of interections 
from viewpoint of orthogonal basis formed by 
eigen-states of unperturbed, i.e. interactionless,  
Hamiltonian \,$H_0 $\, with eigen-values  
\,$\ef_1$\, and \,$\ef_2$\,, and \,$\ef_{21}=\ef_2-\ef_1$\,.     
The GR makes transitions' probabilities 
proportional to time of their expectation 
and thus encourages their treatment in the spirit 
of Marcovian stochastic processes \cite{vk}.   
At that one ignores fluctuations  
of ``unperturbed'' energy (UE) \,$\ef$\, of states 
what are passed in transitions' sequences. 

Meanwhile, the uncertainty principle based estimate 
of steps of these fluctuations, 
\,$\ef_{21}\sim \h/\tau$\,, 
not at all points at their ``vanishing smallness'',  
for their mean square 
\[  
2 \sum_2\,[1 -\cos\,(\tau\ef_{21}/\h)] 
\, |\Phi_{21} |^2 \,\, \,\,  
\]  
has non-zero limit t \,$\tau\rightarrow\infty$\,
and even may be infinitely large. 
This circumstance prompts tbat UE fluctuations 
neglected by kinetics in fact are flicker ones 
(i.e. possessing time-non-integrable correlations), 
which agrees \cite{eph,1803} 
with other results on statistics 
of mechanical thermal motion \cite{1207}-\cite{gs}. 

In reality, of course, that are fluctuations 
in transitions' probabilities and rates 
of relaxation, friction, dissipation, diffusion 
and any other irreversible processes caused by interactions.

Let us show how one can reveal such fluctuations, 
taking in mind interactions between a ``small subsystem'' 
and large ``thermostat'', so that  \,$H_0 \hm = E(V) + H_{th}$\,, 
for instance, 
between ``Brownian'' particle (BP) with \,$E(V)=MV^2/2$\, 
and thermodynamically equilibrium medium.

2.\, In statistical mechanics all interactions are governed 
by the von Neumann equation for system's density matrix (DM) 
 \,$\vro$\,,  
 while kinetics deals with DM's diagonal, 
 that is distribution \,$\rho$\, of probabilities \,$\rho_1 =\vro_{11}$\, 
 of the ``unperturbed'' states. 
If at \,$t=0$\, all non-diagonal MEs were zeros, 
then later     
\begin{eqnarray} \label{de} 
\frac {\pa\rho(t) }{\pa t}  = \int_0^t \qof(t-t^\prime)\,  
\rho(t^\prime) \,dt^\prime \,\, , \,\,\,  
\end{eqnarray} 
where operator \,$\qof$\, is presented by formula  \cite{1803} 
which for a weak interaction (in the framework of second order of NPT) 
reduces to 
\begin{eqnarray} \label{qo} 
 (\qof(\tau)\,\rho)_1  =
 \sum_2 \frac {2|\Phi_{12}|^2} {\h^2}\, 
 \cos \frac {\tau\ef_{12}}{\h}\,\, [\,\rho_2-\rho_1]  \,\, . \,\,  
\end{eqnarray} 
Using the Laplace transformation and marking 
its resultants and related objects with tilde, we have 
\begin{eqnarray} \label{pl} 
\widetilde{\rho}(z) = [ \,z - \qtf_z \, ]^{-1} \,\rho(0) \,\,   \,\,\,  
\end{eqnarray} 
with operator \,$\qtf_z $\, acting by formula    
\begin{eqnarray} \label{lqo}  
(\qtf_z  \,f)_1  = \sum_2 \, \frac {2\pi}\hbar \, |\Phi_{12}|^2 \,
\Delta_z(\ef_{12})\, [\,f_2 - f_1] \,\, , \,\,\, 
\end{eqnarray}
where\, 
\[
\Delta_z( \ep)= \frac 1\pi \, \frac {\h z}{h^2 z^2 +\ep^2}\,\, .  
\]

3.\, Factor   \,$\Delta_z(\ep)$\, determines 
\,$\qtf_z $\,'s properties at small \,$z\rightarrow +0$\, 
and hence \,$\rho(t)$\,'s behavior  at large time.  
For the first look, \,$\Delta_{z\rightarrow +0\,}(\ef_{12})$\, 
must do the same work as the delta-function 
\,$\delta(\ef_{12})$\, in GR and allows 
to write  \,$\qtf_{+0}$\, in (\ref{pl}) in place of \,$\qtf_{z}$\,, 
with \,$\qtf_{+0}$\, being the Marcovian probability evolution 
generator from usual kinetics. 
But a truth is more complex. 
If distribution  \,$f$\, in (\ref{lqo})  
is only depending on UE of states, i.e. \,$f_1\hm =f(\ef_1)$\,, 
then  \,$\qtf_{+0}\,f =0$\,, 
that is GR artificially ``pins'' UE to be constant. 
In fact, however, 
\[
 (\qtf_z  \,f)_1  \,\rightarrow \, z \sum_2 \, \frac {2\pi}\hbar \, |\Phi_{12}|^2 \,
\frac {f(\ef_2) - f(\ef_1)}  
{(\ef_2-\ef_1)^2} \,\propto\, z \,\,
\]
(treating the sun in the sense of principal value),
thus showing 
that during a finite time UE always changes 
by a finite value, although in a slow way. 

Consequently, it will be more right to write
\begin{eqnarray} \label{raz}  
\qtf_z \,\rightarrow\, \qtf_{+0} \, +\, \opi \qtf_z \opi \,\, , \,\,\, 
\end{eqnarray}
where \,$\opi$\, is operator of projecting onto 
functional space of quasi-equilibrium distributions, 
that is ones uniform on any constant UE hyper-surface:\, 
\[
(\opi f)_1 \hm = \mcn^{-1}(\ef_1) \sum_2\,\delta(\ef_1 -\ef_2)\,f_2\,\,,   
\]
where\, \,$\mcn(\ef)\hm = \sum_1\,\delta(\ef -\ef_1)\,$\, is  
density of states with given UE. 
The first term of (\ref{raz}) is responsible 
for fast relaxation to quasi-equilibrium, 
while second term represents slow relaxation, 
like diffusion, over UE axis. 
Because of the latter, 
if starting from \,$\ef_0$\,, UE with time achieves 
distribution,
 \,$W(t,\ef) \hm =\sum_1\, \delta(\ef-\ef_1)\,\rho_1(t)$\,, 
 which ``freezes'', or ``solidifies'', at 
\[
W(\infty,\ef) = \mcn(\ef)\, 
\lim_{z\rightarrow 0}\,z\,\widetilde{\rho}(z) = 
 \frac 1{1 + \widehat{\chi}} \,\,\delta(\ef\! -\! \ef_0)\,  \, , \,  
\]
where operator \,$\widehat{\chi}$\, acts according to   
\[    
\widehat{\chi} \,f(\ef) = 
-\, \frac \h{\pi} \int 
\frac {G(\ef,\ef\pr)} { \mcn(\ef\pr)}\, 
\frac {f(\ef\pr) \! -\! f(\ef)} 
 {(\ef\pr \! -\! \ef)^2}\,d\ef\pr \,    
\] 
with  density of transitions between different UEs 
\[
G(\ef,\ef\pr) = \frac {2\pi}\h  
\sum_{1,2} \delta(\ef_{1}\! -\!\ef)\, |\Phi_{12}|^2 \, 
\delta(\ef_{2}\! -\! \ef\pr) \, \, . \, 
\]
Such the freezing means that UE's fluctuations 
include infinitely long living correlations, 
even with initial conditions,
that is \cite{1803} 
these are flicker fluctuations.

4.\, To consider influence of these fluctuations onto 
the small subsystem, let it be BP, 
we will exploit characteristic function (ChF) 
of integral \,$R(t)=\int _{0}^t V(t\pr)\,dt\pr$\,, 
i.e. BP's path passed during the observation time. 
We define this ChF  \cite{izm} basing on the orrespondence principle, 
in analogy with classical statistical mechanics, as 
\begin{eqnarray} \nonumber 
\Xi(t,ik)=\langle e^{\,ik R(t)}\rangle \,=  
\Tr\, e^{\,t\,(\lof +ik \jof_V)}\, \vro(0) \,\, , \,\,\,\,
\end{eqnarray}
where  \,$\lof$\, and \,$\jof_{V}$\, are quantum Liouville super-operator 
and Jordan super-operator of symmetrized multiplication, respectively, 
 $\jof_{V}\,A\,\hm \equiv (VA + AV)/2$\,.   
Then, quite similarly to derivation \cite{1803} of (\ref{de})-(\ref{qo})  
and  (\ref{pl}), in the second-order NPT one finds   
\begin{eqnarray} \nonumber 
\frac {\pa \Xi(t,ik) }{\pa t}  = ikV\,\Xi(t,ik) + 
\int_0^t \qof(t-t^\prime,ik)\,  
\Xi(t\pr,ik)  \,dt^\prime \,\,   
\end{eqnarray}  
with operator \,$\qof(\tau,ik) $\, which differs from above \,$\qof(\tau) $\, by 
product \,$ \cos{(\tau\ef_{12}/\h)}\,\exp{[ik\,(V_1+V_2)\,\tau/2]} $\, 
in place of cosine \,$\cos{(\tau\ef_{12}/\h)}$\,, 
with \,$ V_{1,\,2}$\, denoting BP's velocity values 
in \,$H_0$\,'s eigen-states \,$1,\,2$\,. 
Thus 
\begin{eqnarray} \label{pl+} 
\widetilde{\Xi}(z,ik) = \Tr\, [\, z\! -\! ik V  -   
 \qtf_z(ik) ]^{-1} \rho(0) \,\, , \,\,\,  
\end{eqnarray} 
where operator \,$\qtf_z(ik)$\, differs from \,$\qtf_z$\, 
by replacement  
\begin{eqnarray} \label{zam} 
 \Delta_z(\ef_{12})\, \Rightarrow\, 
 \Delta_{z-ik\,(V_1+V_2)/2\,} (\ef_{12})\,\, \,\,\,\,  
\end{eqnarray} 
in the sum inside it  
(so that \,$\qtf_z(0) \hm = \qtf_z$\,). 
This is expression of interference between BP-medium interaction 
and BP's motion.

For description of this interference let us write    
\[
 \qtf_z(ik) - \qtf_z(0) \hm \equiv ik \,\vtf_z(ik)\,\,  \, 
\] 
and introduce operator of (temp of) BP's diffusion:   
\begin{eqnarray} \label{dtf} 
\dtf_z(ik) = \opi\, (V+\vtf_z(ik))\opi\pr \, \,\times \,\,\,\,\, 
\\ \nonumber \, \times\,  
\{z -ikV - \opi\pr\qtf_z(ik) \opi\pr\}^{-1} 
\opi\pr (V+\vtf_z(ik))\opi \,\,  \,  
\end{eqnarray} 
with \,$\opi\pr \equiv 1-\opi$\,.
The ChF reduces to it,  
when initial distribution is (quasi-) equilibrium,
i.e. \,$\opi \rho(0)\hm = \rho(0)$\,, 
for example, \,$\rho(0)\hm =\delta(\ef-\ef_0)/\mcn(\ef_0)$\,. 
Then   
\[
\widetilde{\Xi}(z,ik) \rightarrow \, \Tr\, 
\{ z\! + k^2 \dtf(z,ik) +O(z^2) \}^{-1}\,\opi\rho(0) \,\,  
\]
 under \,$z\rightarrow 0$\, with \,$k$\,-independent \,$O(z^2)$\,. 
If, in addition, the interaction is uniform 
(invariant in respect to shifts) on UE axis,
so that the above quantities 
\,$G(\ef,\ef\pr)/\mcn(\ef\pr)$\, and \,$W(t,\ef)$\, 
in fact depend only on differences \,$\ef\! -\!\ef\pr$\, 
and \,$\ef\! -\!\ef_0$\,, 
then it is not hard to guess and prove that 
\begin{eqnarray} \nonumber   
\widetilde{\Xi}(z,ik)  \rightarrow   
\frac 1{ z\! +\! k^2 \Tr\,\dtf_z(ik) \, \rho(0) } \equiv 
 \frac 1{ z\! + k^2 \df(z,ik) } \,\,  
\end{eqnarray} 
with \,$ \df(z,ik) $\, insensible to (quas-equilibrium)  \,$\rho(0)$\,.

5.\, Now we have to discuss function\, 
\,$ \df(z,ik)  \hm = \Tr\,\dtf_z(ik) \, \rho(0)$\,.
Operator \,$\opi\pr\qtf_z\opi\pr$\, in  (\ref{dtf}) 
represents fast relaxation of BP's velocity distribution to 
equilibrium one (balanced with medium). 
Presuming \,$k$\, and \,$z$\, to be infinitesimally small, 
it seems much reasonable to approximate contents of the braces 
in (\ref{dtf}) by \,$-\qtf_{+0}$\, 
and the latter, for simplicity, 
by characteristic eigen-value of \,$-\qtf_{+0}$\,, 
i.e. characteristic rate (inverse time) 
\,$g_0=1/\tau_0$\, 
of the velocity relaxation. 
This roughening helps us to focus at much more 
important role of operator  \,$\vtf_z(ik)$\, 
which mixes UE fluctuations to BP's wandering.  
With taking into account that
\,$\Tr\,\vtf_z(ik)\,\dots\hm =0$\, we have    
\begin{eqnarray}  \label{df0} 
\df(z,ik)  \rightarrow\,\tau_0\,
\Tr\,V\, [V + \vtf_z(ik)\,\opi ]\,\rho(0) \,\, , \,   
\end{eqnarray} 
where action of \,$\vtf_z(ik)$\, can be presented, 
in view of obvious symmetries of interaction MEs, by formula   
\begin{eqnarray}  \nonumber  
(\vtf_z(ik)  \,f)_1  = \sum_{2,\pm} \,  
\frac { V_{12}\, |\Phi_{12}|^2 \,[\,f_2 - f_1]} 
{(\h z\pm i\ef_{12})^2 + (\h k V_{12})^2}  \,\,  \,\,\, 
\end{eqnarray}
with\, \,$V_{12} =(V_1+V_2)/2$\,.

Next, it will be comfortable to separate BP's velocity 
from the full system states' indices \,$1,\,2,\,\dots$\, 
and instead of \,$V_1,\,V_2,\,\dots$\, 
write \,$V,\, V\pr,\,\dots$\, 
while giving ciphers \,$1,\,2,\,\dots$\, 
to medium's states (eigen-states of \,$H_{th}$\,). 
Then  \,$\rho_1$\, turns to \,$\rho_{V1}$\, and 
\begin{eqnarray}  \nonumber 
 \sum_1 \rho_{V1}(0) = \sum_1 \,
 \frac {\delta(\Sigma_1\! +\! E(V)\!-\!\ef_0)}{\mcn(\ef_0)} = 
 W_0(V) \, \equiv \, 
 \\ \nonumber \, \equiv\, 
  \frac {\mcn_{th}(\ef_0\!-\!E(V))} {\mcn(\ef_0)} = 
 \frac {\mcn_{th}(\ef_0)} {\mcn(\ef_0)}  \, \exp{[-E(V)/T]}\,\, , \, 
\end{eqnarray}
where \,$\Sigma_1$\, are energies of medium states,  
\,$\mcn_{th} (\Sigma)$\, is their density, 
\,$T \hm = [d\ln{\mcn_{th}(\ef)}/d\ef] ^{-1}$\, is temperature 
of micro-canonical ensemble  \cite{ter} of medium states,
and \,$W_0(V) $\, is Maxwellian equilibrium distribution of BP's velocity.
Using these designations and mentioned interaction uniformity and symmetries, 
we can write 
\begin{eqnarray}  \label{df} 
\df(z,ik) \, \rightarrow\,  D_0 + \tau_0
\iint \! \int \frac {\h\,V\,(V+V\pr)} {8\pi} \,   
\,\times \,\,\,\,\,\,  \\ \nonumber \times \, \sum_\pm\, 
\frac { W(V,V\pr;\ep) - W(V\pr,V;-\ep) } 
{(\h z\pm i\ep)^2 + (\h k\,(V+V\pr)/2)^2} \,\, \,\,  
\end{eqnarray}
with integrals over velocities \,$V,\,V\pr$\, 
and over UE deviation from conservation, \,$\ep$\,, 
with function   
\begin{eqnarray}    \label{fw} 
W(V,V\pr;\ep) = \frac {2\pi}{\h} \sum_{1,2} \,
\delta(\ef_0+\ep\! -\! \Sigma_{1} \! -\! E(V))\, 
\,\times \,\,\,\,\,\,  \\ \nonumber \times \,
|\Phi_{V1\,V\pr 2}|^2 \,\, \frac 
{\delta(\ef_0 \! -\! \Sigma_2\! -\! E(V\pr)) } 
 {\mcn(\ef_0)} \,\,  \,    
\end{eqnarray}
and with\, \,$D_0\hm = \tau_0 \int V^2\,W_0(V) \,dV$\,. 
Simultaneously, in terms of only velocity of BP 
(under ``manually kept'' constant UE value), 
operator \,$\qtf_{+0}$\,  becomes   
\[
 \qtf_{+0}\,f(V) =\int W(V,V\pr;0)\, 
 \left[ \frac { f(V\pr) }{W_0(V\pr)} - 
 \frac {f(V)}{ W_0(V)} \right]\,dV\pr\,\, . \, 
\]

6.\, Now, consider the integrals in (\ref{df}), 
for brevity dealing with \,$V$\, like velocity of one-dimensional motion 
along \,$k$\,'s direction. 
Notice that function  (\ref{fw}) by very its definition 
is proportional to (relative) density of medium states 
at lowest of its  ``left'' and  ``right'' energies, 
\,$\Sigma_1 \hm = \ef_0+\ep\! -\! E(V)$\, 
and \,$\Sigma_2\hm = \ef_0\! -\! E(V\pr)$\,.  
It means that 
\begin{eqnarray}    \label{str} 
W(V,V\pr;\ep)  \,=\, w(V,V\pr; |E-\ep -E\pr|)\, 
\times  \,\,\,\,\,  \\ \nonumber  \,\times\,\, 
\exp{\left(-\frac {E-\ep +E\pr+\, |E-\ep -E\pr|}{2T} \right)} \,\,  \,    
\end{eqnarray}
with \,$E=E(V)$\, and \,$E\pr =E(V\pr)$\,, 
where, again due to the definition  (\ref{fw}), symmetry 
\,$ w(V\pr,V;\sigma) \hm = w(V,V\pr \sigma)$\, 
takes place, 
and factor  
 \,$\sigma \hm \equiv |E-\ep -E\pr| \hm = |\Sigma_1 -\Sigma_2|$\, 
 represents 
 energy donated or obtained by medium during a transion. 
 We also took into account  that multiplier 
  \,$ w(V\pr,V;\dots) $\, characterizes excitations of medium in itself, 
  therefore  it may involve the energy  ``discrepancy'' (deviation), 
  \,$\ep$\,, only just through medium energy change \,$\sigma$\,. 
Besides, if medium along with its interactions  
is spatially uniform and isotropic, then 
 \,$ w(\cdot) $\,  must have only two scalar arguments:\, 
 \,$w(V,V\pr;\sigma)\hm \Rightarrow w(|V-V\pr|;\sigma)$\,. 
Consequenntly,   
\begin{eqnarray}  \label{dfq} 
\df(z,ik) \, \rightarrow\,  D_0 + \tau_0
\iint \! \int \frac {\h\,(E-E\pr)} {4\pi M} \,   
\,\times \,\,\,\,\,\,  \\ \nonumber \times \, \sum_\pm\, 
\frac { \sinh{(\ep/2T)}\, F(V,V\pr;|E-\ep-E\pr|)}   
{(\h z\pm i\ep)^2 + (\h k\,(V+V\pr)/2)^2} \,\, , \,\,  
\end{eqnarray}
where
\[
F(V,V\pr;\sigma) \equiv\, w(|V\!-\! V\pr|;\sigma)\, 
 \exp{[-(E\! +\! E\pr \! + \sigma)/2T]} \,\,
 \]
and, evidently, only odd in respect 
at once to \,$\ep$\, and  \,$E-E\pr$\, component of \,$F(V,V\pr;|E-\ep-E\pr|) $\, 
in fact contrbutes to the integrals. 

Formula (\ref{dfq}) is main result of our communication. 
In its rest we point out some of consequences from (\ref{dfq}).

7.\, Consider expansion  
\[
\df(z,ik) = \df_0(z) + \df_2(z) \, (ik)^2 + \dots\,\, . 
\] 
First of all we want to know about behavior of coefficient 
\,$\df_2(z) $\, when \/$z\rightarrow +0$\,, 
since it is coonected to long-time asymptotics 
of fourth-order cumulant of the BP's path: 
\begin{eqnarray} \nonumber 
\int_0^\infty e^{-zt} \, 
[\,\langle R^4(t)\rangle - 3 \langle R^2(t)\rangle ^2\,]\,dt \rightarrow  
\frac  {\df _2(z)} {z^2 }  \,\, , \,\,
\end{eqnarray}
and thus says about large-scale deviations of BP's wandering 
statistics from the Gaussian one. 

From (\ref{dfq}) it follows that  
\begin{eqnarray} \nonumber 
\df_2(z)  =  \left[ 1 + \frac 52\, z^2\,\frac {\pa}{\pa z^2} + 
\frac 56\, z^4\,\left(\frac {\pa}{\pa z^2} \right)^2\right]\,\,  
\, \times \,\, \\  \nonumber \times \,\, 
\frac {\tau_0 \h^3}{4\pi M} \int \,    
\frac  {\sinh{(\ep/2T)}\, I_2(\ep) } 
{(\ep^2+\h^2z^2)^2}\,\, d\ep\,\,    
\end{eqnarray}
with second of velocity integrals
\[
 I_{2n}(\ep) = \iint_{V,V\pr} v^{2n}\,  
 (E\! - \! E\pr) \, F(V,V\pr;|E\! -\! E\pr \! -\! \ep|)\,\, ,  
\]  
where\, \,$v \equiv (V\! +\! V\pr)/2$\,.   
All they are odd functions of \,$\ep$\,.  
It is visible from here that if 
\,$I_2(\ep\rightarrow 0) \propto \ep$\,, 
then 
\begin{eqnarray}   \label{d2a} 
\df_2(z\rightarrow 0) \, \rightarrow \, \frac 1z\,\, 
\frac {3 \tau_0 \h^2}{64\, MT} \, 
\left[\frac {d I_2(\ep) } {d\ep} \right]_{\ep =0}  + \,\texttt{const}\,\, . \,     
\end{eqnarray}
This means that absolute value of 
the fourth-order path cumulant grows with observation time 
proportionally to its square. 
In other words, non-Gaussianity of the wandering 
does not decrease with time at all. 

At that, seemingly, both positive 
and negative signs of the non-Gaussianity are possible, 
dependently on sign of  \,$[dI_2(\ep)/d\ep]_{\ep =0}$\,. 
Positive case allows natural interpretation 
as the result of smooth flicker fluctuations 
of  BP's diffusivity (diffusion coefficient)  \cite{bk}-\cite{157}. 
Then their effective correlation function   
\,$C_D(\tau)$\, (asymptotically) is determined by relation\,   
\[
\int_0^\infty C_D(\tau)\,\exp{(-z\tau)}\, d\tau 
\hm \rightarrow \df_2(z)\,\, . \, 
\]
So, asymptotics (\ref{d2a}) implies 
\,$C_D(\infty)=$\,const\,$> 0$\,, 
that is, verbally, ``quasi-static'' fluctuations.

Negative sign in (\ref{d2a}), of course, 
also reflects flicker fluctuations of BP's wandering, 
but  of some different type, 
probably discontinuos like ``telegraph signals''. 
 
In particular, both signs can realize 
when   variety of medium's energy  changes  \,$\sigma$\, 
(irradiated or absorbed quanta) possible  in one transition 
consists of single number 
(like in the simple phonon medium in \cite{eph}):\,
 \,$w(|u|;\sigma) \hm = w(|u|)\, \delta(\sigma- q(|u|))$\,, 
 where \,$u \equiv V\! -\! V\pr$\,. 
 At that, positive and negative contributions 
 to the fourth cumulant are coming 
 mainly from quanta \,$q(|u|)$\,
 smaller and greater than medium temperature, respectively.

8.\, Complete enough analysis of \,$\df(z,ik) $\, 
as a whole we leave for the future. 
Just here we have only to comment  
BP's diffusivity (diffusio coefficient) as such, i.e. 
  \,$\df_2(0)\hm =\df(0,0) $\,, which is given by  
\begin{eqnarray}   \label{d0a} 
\df_0(z\rightarrow 0)  \rightarrow  D_0 - 
\frac {\tau_0 \h}{2\pi M} 
\int \frac {\sinh(\ep/2T)\, I_0(\ep) } {\ep^2} \, d\ep\,\, . \,\,\,      
\end{eqnarray}
There,  second (integral) term also, like (\ref{d2a}), 
is able for any sign, that is it can 
both increase and decrease 
the seed value \,$D_0$\, given by the usual kinetics. 

It is necessary to underline insensitivity of the second term 
to degree of weakness of the interaction:\, 
since 
 \,$w(\cdot) \hm \propto g_0=1/\tau_0$\, 
by physical meaning as well as formal definitions of these objects, 
the product  \,$\tau_0 I_{0}(\ep)$\, is indifferent to \,$\tau_0$\,. 
At the same time \,$D_0 \propto \tau_0$\,, 
therefore, in case of positivity of integral in (\ref{d0a}), 
too strong  diminution of \,$\tau_0$\, would turn 
\,$\df_0(0) $\, to zero or less. 
But, clearly, it would be mere artifact of 
too unwary application of low-order NPT 
(in strict all-orders NPT \,$\tau_0$\, hardly can turn to zero).

9.\,  In conclusion, one more principal remark.  
Our above reduction of formalism from full micro-states 
to the pair of variables \,$\ef,\, V$\, 
has cut off possibilities to include effects of medium's memory. 
But, instead, we have concentrated on most 
fundamental effects of unavoidable time-energy uncertainty 
in real-life, finite-duration, interactions and observations. 
We hope that it is a noticable progress 
in quantum microscopic theory of ``pure'' 
flicker (1/f-) noise, that is one which, - 
as already was shown  \cite{lufn,tmf,gs} 
in classical statistical mechanics, - 
is created by interaction even with such 
memoryless media as ideal gas.

\,\,\, 



\,\,

\vfill\eject


\begin{thebibliography}{99} 

\bibitem{ll3}   
Landau L D, Lifshitz E M, {\it Quantum mechanics} (Pergamon Press, 1977)   

\bibitem{vk}
van Kampen N G, {\it Stochastic processes in physics and chemistry} 
(North-Holland, 1992)  

\bibitem{hj}  Haug H, Jauho A-P. {\it  Quantum kinetics in transport 
and optics of semiconductors}  (Springer-Verlag, 2007) 


\bibitem{eph}  Kuzovlev Yu E, arXiv:1704.01542 

\bibitem{1803}  Kuzovlev Yu E, arXiv:1803.09250 


\bibitem{1207} Kuzovlev Yu E, arXiv:1207.0058 

\bibitem{lufn}
Kuzovlev Yu E  {\it Physics - Uspekhi}  {\bf  58} (7)  719  (2015); 
arXiv:1504.05859  
 
\bibitem{i1} 
Kuzovlev Yu E,  Sov. Phys. - JETP\,  {\bf 67} (12) 2469 (1988); 
arXiv:0907.3475 

\bibitem{tmf}
Kuzovlev Yu E,  {\it Theor. Math. Phys.} {\bf 160} 1301 (2009); 
arXiv:0908.0274 

\bibitem{tof} Kuzovlev Yu E, arXiv:1008.4376 

\bibitem{pufn}
Bochkov G N, Kuzovlev Yu E,  {\it Physics - Uspekhi}  {\bf  56} (6) 590 (2013); 
arXiv:1208.1202   

\bibitem{gs} 
Kuzovlev Yu E,  {\it JETP Letters} {\bf 103} (4) 234  (2016); 
arXiv:1512.04835 


\bibitem{izm} Kuzovlev Yu E, arXiv:cond-mat/0501630

\bibitem{ter}
Terletskii Ya P,  {\it Statistical physics} (Moscow: Vysshaya Shkola, 1994) 
(in Russian) 


\bibitem{bk}
 Kuzovlev Yu E, Bochkov G N,  {\it Radiophysics Quant. Electronics} 
 {\bf 26} (3) 228 (1983); {\bf 27}  811 (1984)  

\bibitem{ufn}
Bochkov G N, Kuzovlev Yu E, Sov. Phys. - Uspekhi\, {\bf  26} 829 (1983)   

\bibitem{99}  Kuzovlev Yu E, arXiv:cond-mat/9903350  

\bibitem{157} Kuzovlev Yu E, arXiv:1211.4167 



\end{thebibliography}
\end{document}